# Excelsior: Bringing the Benefits of Modularisation to Excel

*Jocelyn Paine*

*www.j-paine.org/* and *www.spreadsheet-factory.com/*

**ABSTRACT**

*Excel lacks features for modular design. Had it such features, as do most programming languages, they would save time, avoid unneeded programming, make mistakes less likely, make code-control easier, help organisations adopt a uniform house style, and open business opportunities in buying and selling spreadsheet modules. I present Excelsior, a system for bringing these benefits to Excel.*

## 1 INTRODUCTION

There are two ways of constructing a software design. One is to make it so simple that there are obviously no deficiencies; the other is to make it so complicated that there are no obvious deficiencies. The first method is far more difficult. [Hoare, 1981]

The only way to write complex software that won't fall on its face is to build it out of simple modules connected by well-defined interfaces, so that most problems are local and you can have some hope of fixing or optimizing a part without breaking the whole. [Raymond, 2003]

You can model the value of modularity in a decision-theoretic way by using the Black-Scholes equation. Modularity increases your set of options for modifying a system, and it's a basic result of options theory that having a portfolio of individual options is more valuable than having an option on a portfolio. (Intuitively, a program whose components you can selectively upgrade is more valuable than one which is a monolith which must be upgraded in a big bang.) But to use that result, it must actually be the case that your modules are modular - and that's where you need abstraction as a semantic property. [Krishnaswami, 2005]

If you want to avoid having a missed bug affect an arbitrarily large proportion of the system, the answer is compartmentalization (more modularity), with effective enforcement of the boundaries between compartments. These mechanisms are able to prevent a missed bug from affecting an arbitrarily large proportion of the system. [Sitaker, 1998]

Spreadsheets lack the most fundamental mechanism that we use to control complexity: the ability to define re-usable abstractions. They deny to end-user programmers the most powerful weapon in our armory. Can you imagine programming in C without procedures, however clever the editor's copy-and-paste technology? [Peyton Jones et. al., 2003]

This paper describes research on providing Excel with features for modular design, providing the advantages and avoiding the complaint voiced above. Section 2 explains modular design and its benefits, and should interest spreadsheet users and managers not lucky enough to have encountered these already. Section 3 introduces the research plan and how it fits with the rest of the paper. Sections 4 and 5 describe a simple mathematical treatment of spreadsheets and its use for adding modularity, with examples. Section 6 describes Excelsior, a programming language based on these principles. Section 7 briefly explains how this work could be continued and fitted to Excel. Section 8 briefly explains how Excelsior relates to, and can be used from, the Prolog programming language. Section 10 describes four applications.

## 2 MODULAR DESIGN AND ITS BENEFITS

*Modular design* is making an object - be it car, television, disc drive, or spreadsheet - from parts or *modules* which can be understood each on its own, not needing knowledge of other parts with which we might join it. The idea is to think about modules' insides separately from their outsides. Speaking generally – this applies as much to, say, mechanical engineering as to software - a module makes a "contract" that it will, given specified inputs, generate specified outputs. All the user need know is where to connect the inputs and outputs, and how they are related. The rest of the module's workings are up to its implementor, as long as he or she honours the contract.

### 2.1 Top-down design - plans for understanding, writing, testing, and documenting spreadsheets

*Top-down design* means thinking in terms of a hierarchy of modules. When fault-finding a hi-fi, we don't break into the chips: we read their specifications, ensure we understand why these components are connected as they are, and test that each is in fact receiving the right inputs. Thus, the structure of something – including a spreadsheet - built from modules gives us a plan to follow when trying to understand it.

In the same way, top-down design gives us a plan to follow when writing, testing and documenting new ones. Code one module at a time; test each thoroughly, then treat as a reliable component; reuse modules wherever possible. Most programmers know this: I emphasise it because some with only spreadsheet experience will not.

### 2.2 Localisation of concerns - easy updating, less work, fewer mistakes, a uniform house style

When we want the same calculation in different parts of a program, or in different programs, we could code it anew each time. Or we could code it once as a module, then insert that wherever we need it. This is the *localisation of concerns*: all the code concerned with a particular calculation is in a single clearly defined location.

Localisation of concerns saves time: instead of constantly redesigning, recoding, retesting, and redocumenting some calculation, the programmer just inserts a module from a library. It reduces risks: the less code, the fewer mistakes. And it saves work when updating code. Suppose we have many programs, all needing the same calculation. And suppose we want to extend the range of inputs the calculation can handle, or we find in it a mistake. If all programs use the same module to do the calculation, that's the only thing we need fix. But if the programmer has coded the calculation anew for each program, they need fix it as many times as there are programs; finding all these programs becomes difficult; new mistakes become more likely. Localisation of concerns also helps organisations achieve a uniform house style.

### 2.3 Interchangeability of parts and suppliers

If all that's important about a module is its specification and external connections, we can replace it by any other module having identical specification and external connections, even if the workings are different. This is *interchangeability of parts*. In spreadsheeting, we could replace one module by another that's faster or more precise; that has been enhanced to handle complex numbers or Euros; that is cheaper, has better customer support, or a vendor less likely to go bust.

### 2.4 Code control

Being able to divide spreadsheets into modules would also aid code control, since we can track mistakes and alterations to each part separately. However, most code-control systems want source code as text, not as spreadsheets. The software I describe here makes this possible by enabling spreadsheets and spreadsheet modules to be saved as text.

### 2.5 Business opportunities

Modules for spreadsheets would create business opportunities in selling spreadsheet parts which others can use. Conversely, businesses could save effort and reduce bugs by out-sourcing to experts. It's like wiring up a mass of circuitry from scratch, versus buying tested and guaranteed circuit boards, each with predefined connections, operating specifications, guarantee, and customer support.

### 3 RESEARCH PLAN

The research described here breaks down into the following steps:

**1.** Finding a mathematical representation for spreadsheets. This is explained in the next section.

**2.** Defining functions which act on objects so represented, combining and extracting parts or modules, and bearing in mind the points about modularity made in Sections 2 and 7. These functions are described in Section 5.

**3.** Designing a programming language to be as close as possible to these. This is Excelsior, explained in Section 6.

**4.** Excelsior is implemented in Prolog. There are sound reasons for not inventing yet another programming language when so many already exist. On the other hand, if we provide instead just a library for use from Prolog, we are bound by Prolog's syntax and semantics, inconvenient for many users. This stage, therefore, looked at the pros and cons of implementing a Prolog interface to Excelsior, described in Section 8.

**5.** By now, Excelsior had almost become a practical language for handling spreadsheets, usable as a "scripting language" by programmers accustomed to such as Perl, Python, and Rexx. In this stage, therefore, were added features likely to increase its usefulness, such as formula editing. These are explained in Section 6.

**6.** The operators of Section 4.2 - can transform spreadsheets to be easier to understand. However, they need information about the spreadsheet author's intentions. Trying to discover these is the "structure discovery" of Section 4.6. In this stage, therefore, were added other features useful to structure discovery.

**7.** Excelsior is not for the typical Excel user; but it is a good foundation for a modularisation tool that could be used by them. Designing this, bearing in mind users' skills and expectations, is the next stage of research. This is explained in Section 7.

**4 SPREADSHEETS AS MATHEMATICAL OBJECTS**

By treating spreadsheets as mathematical objects, this section defines operators for constructing new spreadsheet parts, splitting existing ones into parts, abstracting them, and joining them together. Unlike with other programming languages, these must take into account spreadsheets' visual aspect - layout - as well as their content. Mathematicians may note that the work was inspired by category theory [Goguen, 1991] (useful in generalising the notion of "putting together"), sheaf semantics [Goguen, 1992], horizontal and vertical module composition [Goguen, 1996], and abstraction [Tennant, 1981].

**4.1 Spreadsheets are sets of equations**

Spreadsheets are, in essence, sets of equations. Here's an example; these equations could have come from a simple accounting spreadsheet:

```
A2 = 2000
A3 = 2001
B2 = 1492
B3 = 1560
C2 = 971
C3 = 1803
D2 = C2 - B2
D3 = C3 - B3
```

**4.2 Improving intelligibility by transforming equations**

Transforming a spreadsheet's equations can make them clearer. Suppose for example that the author of the spreadsheet above meant column B to be expenses, C to be sales, and D to be profits. Rewriting the equations accordingly makes them easier to understand:

```
Year[2000] = 2000
Year[2001] = 2001
Sales[2000] = 971 etc.
Profit[2000] = Sales[2000] - Expenses[2000]
Profit[2001] = Sales[2001] - Expenses[2001]
Layout Year[2000:] as A2 downwards
Layout Expenses[2000:] as B2 downwards
Layout Sales[2000:] as C2 downwards
Layout Profit[2000:] as D2 downwards
```

**4.3 Compiling spreadsheets from specifications**

Transforming equations in the other direction lets us specify a spreadsheet's calculations separately from its layout, and in a way that uses meaningful identifiers and cell groupings. For example, we could transform the second example to the first by replacing `Year[2000]` by `A2`, `Year[2001]` by `A3`, and so on as directed by the layout statements. This is *compiling* equations to a spreadsheet.

**4.4 Separating calculation from layout**

Compiling is powerful, because by varying the layouts, we can map the same equations to a spreadsheet in many different ways. Thus we could compile the same three-dimensional table to vertically stacked 2-d cross-sections in one spreadsheet, to a horizontal sequence of cross-sections in another, and to one slice per worksheet in a third. It would also, for example, be equally easy to change the layouts so data runs from right to left, to accommodate Hebrew and Arabic readers.

**4.5 Decompiling spreadsheets into specifications**

The opposite to compiling is *decompiling*. This starts with the spreadsheet and rewrites its equations to be more intelligible, as in Section 4.2.

**4.6 Structure discovery**

Before we can decompile a spreadsheet, we need to know how its author meant its cells to be grouped. For example in the above example, cells `A2` and `A3` were meant to form a table of years. We also need sensible names for them. This - uncovering its author's intentions – is *structure discovery*.

As [Clermont, 2004] and [Hipfl, 2004] explain, we can devise heuristics to guess this implicit structure. To find out which cells belong in the same array, we can seek: regions surrounded by text or blank cells; sequences of identical formulae; ranges passed to SUM and other aggregating functions. To discover the array bounds, we can look for sequences of numbers such as years, and for sequences of common words such as month names. These often act as subscripts to the cells below or on their right. To guess names for the arrays, we can use labels in neighbouring cells. Because Excelsior is a complete programming language, any such heuristic can be coded in it; to simplify this, it provides spreadsheet grammar rules [Paine, 2004(a)].

**5 PRIMITIVES FOR MODULARISING SPREADSHEETS**

I shall use the baby accounting spreadsheet of Section 4.1 as a running example.

**5.1 The 7 operator: combining sets of equations**

Suppose we want to put labels over the columns to say that column A is year, B expenses, C sales, and D profit. Let's write these labels as equations:

```
A1 = "Year"
B1 = "Expenses"
C1 = "Sales"
D1 = "Profit"
```

Then to insert the labels, we need only combine these equations with those defining the account spreadsheet. I shall do this, using the brackets `{` and `}` to enclose sets of equations, and K to combine them. (K forms the union of two sets under the constraint that there cannot be two equations with the same left-hand side but different right-hand sides.). The spreadsheet we want, with accounts and labels, can then be written as:

```
{ A1 = "Year", B1 = "Expenses", C1 = "Sales", D1 = "Profit" } K
{ A2 = 2000, A3 = 2001, B2 = 1492, B3 = 1560,
  C2 = 971, C3 = 1803, D2 = C2-B2, D3 = C3-B3 }
```

We can insert calculations too. Suppose we wanted to calculate the tax - assume it's 33% - on profits and show the result in column E. We can write:

```
{ A2 = 2000, A3 = 2001, B2 = 1492, B3 = 1560,
  C2 = 971, C3 = 1803, D2 = C2-B2, D3 = C3-B3 } K
{ E2 = D2×0.33, E3 = D3×0.33 }
```

So K lets us insert cells that are independent of existing cells (the labels), but also cells that depend on existing cells (these calculations).

### 5.2 The 7 operator: shifting equations

Now suppose we want to insert a column of text on the left of column A. We need to shift existing columns right, which I'll do with K. This takes a set of equations and an (X,Y) pair, and shifts the spreadsheet represented by the equations X places right and Y down:

```
{ D3 = C3-B3, D2 = C2-B2 } K (2,10) =
{ F13 = D13-C13, E12 = D12-C12 }
```

Then if (say) we wanted to insert a column of years as Roman numerals, we could write:

```
{ A2 = "MM" , A3 = "MMI" } K
{ A2 = 2000, A3 = 2001, B2 = 1492, B3 = 1560,
  C2 = 971, C3 = 1803, D2 = C2-B2, D3 = C3-B3 } K (1,0)
```

The K and K operators therefore have two purposes which fit well together: joining parts of spreadsheets visually; and chaining together inputs to outputs.

### 5.3 The `let`: naming sets of equations

In future examples, I shall want to name sets of equations. I do so with `let`:

```
let accounts = { A2 = 2000, A3 = 2001, B2 = 1492, B3 = 1560,
                 C2 = 971, C3 = 1803, D2 = C2-B2, D3 = C3-B3 }.
let tax = { E2 = D2×0.33, E3 = D3×0.33 }.
accounts K tax
```

### 5.4 The `@` operator: extracting subsets from a set of equations

Suppose we want to insert a column in the middle of a spreadsheet. This needs us to split parts out of the spreadsheet. I do so with `@`, which takes a set of equations and a cell range, and returns those equations whose left-hand sides lie within the range. Thus:

```
{ A2 = 2000, A3 = 2001, B2 = 1492, B3 = 1560,
  C2 = 971, C3 = 1803, D2 = C2-B2, D3 = C3-B3 } @ A1:D2 =
{ A2 = 2000, B2 = 1492, C2 = 971, D2 = C2-B2 }
```

We can then insert a column this way:

```
accounts @ A:C K
( accounts @ D:D ) K (1,0) K
{ D2 = E2×0.33, D3 = E3×0.33 }
```

### 5.5 Building one spreadsheet from two

With `@`, we can create a chimera. Suppose we have another spreadsheet `accounts2`, which has one column of expenses data in B. And suppose we want to use this in `accounts`. We can glue the pieces together like this:

```
let expenses2 = accounts2 @ B.
accounts @ A:A K expenses2 K accounts @ C:D
```

See how naturally we now dissect components out of spreadsheets and recombine them.

### 5.6 Using `@` to combine non-contiguous ranges

The `@` operator makes sense with non-contiguous ranges:

```
let expenses2 = accounts2 @ B.
accounts @ (A:A,C:D) K expenses2
```

or even

```
accounts @ (A:A,C:D) K accounts2 @ B
```

### 5.7 The `mapping` operator: changing layouts

When reusing part of a spreadsheet, we may need to change its layout to fit with another spreadsheet. We also need to change layouts when compiling and decompiling. For this, I introduce the `mapping` operator: it takes a set of equations on its left, and a source and a target cell range on its right, and rewrites the equations so that elements in the source are replaced by corresponding elements in the target.

Suppose we have a spreadsheet `accounts3`. Like `accounts` and `accounts2`, this holds expenses data, but running horizontally from A1 to A2. To use this in `accounts`, we need to rotate this row onto B2:B3. Then we can do so with `mapping`:

```
let expenses3 = accounts3 @ A1:A2.
let expenses3_rotated = expenses3 mapping A1:A2 to B2:B3.
accounts @ A:A K expenses3_rotated K accounts @ B:D
```

### 5.8 Arrays in equations

It's essential we can combine intelligible specifications like that of Section 4.2 with spreadsheets. We may want to decompile existing spreadsheets into such specifications and use those as the master code. So we must be able to use arrays in equations. We do so very naturally:

```
let accounts_S = { Year[2000] = 2000, Year[2001] = 2001,
                   Expenses[2000] = 1492, Expenses[2001] = 1560,
                   Sales[2000] = 971, Sales[2001] = 1803
                   Profit[2000] = Sales[2000]-Expenses[2001],
                   Profit[2001] = Sales[2001]-Expenses[2001] }.
```

Similarly, we could write the tax formulae of Section 5.3 as:

```
 let tax_S = { Tax[2000] = Profit[2000]×0.33,
               Tax[2001] = Profit[2001]×0.33 }.
```

**5.9 Worksheets as arrays**

It turned out convenient for us treat worksheets as arrays too. Thus the accounts spreadsheet is actually as shown below, the cell notation being syntactic sugar:

```
  let accounts = { Sheet1[1,2] = 2000, Sheet1[1,3] = 2001,
                   Sheet1[2,2] = 1492, Sheet1[2,3] = 1560,
                   Sheet1[3,2] = 971,  Sheet1[3,3] = 1803
                   Sheet1[4,2] = Sheet1[3,2]-Sheet1[2,2],
                   Sheet1[4,3] = Sheet1[3,3]-Sheet1[2,3] }.
```

**5.10 The × operator: replicating sets of equations**

This replicates an object along a new dimension:

```
  { y[1] = 1 } × 2000:2001 =
  { y[1,2000] = 1, y[1,2001] = 1 }
```

This is a good way to build tables with repeated parts, such as a spreadsheet holding one copy of a chain-store's accounting proforma for every branch of the store.

**5.11 The / operator: dividing or quotienting sets of equations**

The operator / is the converse of multiplication. It takes the quotient of an object, projecting it through its final dimension:

```
  { y[1,2000] = 1, y[1,2001] = 1 } / 2000:2001 =
  { y[1] = 1 }
```

For this to be possible, all the formulae that project onto the same result formula must be equivalent. Variants of this operator will probably be useful in concisely specifying searches for repeated structure when decompiling spreadsheets.

**5.12 The `contents_of` function: code reuse and spreadsheet libraries**

At last, I shall show how to reuse code from Excel files and from files holding sets of equations. To do so, I hall introduce the contents_of function. This takes a string as argument, assuming it to name a file. If the filename ends in .xls, we have an Excel file; if in .exc, we assume it to hold equations written as in this paper.

This gives us libraries and code reuse. Look again at the example from Section 5.3:

```
  let accounts = { A2 = 2000, A3 = 2001, B2 = 1492, B3 = 1560,
                   C2 = 971, C3 = 1803, D2 = C2-B2, D3 = C3-B3 }.
  let tax = { E2 = D2×0.33, E3 = D3×0.33 }.
  accounts K tax
```

Now rewrite it as

```
  let accounts = contents_of( "accounts.xls" ).
  let tax = contents_of( "tax.xls" ).
  accounts K tax
```

In general, we can call `contents_of` anywhere a set of equations is needed. These equations could have been written by hand or pulled from existing spreadsheets. In either case, they could either be "raw" spreadsheet equations using cell names, or equations to be compiled, written using meaningful arrays and names. We have achieved code reuse and modularity.

## 6 EXCELSIOR

Excelsior is an executable equivalent of the last section: the nearest a keyboard can come to the notation. It is *functional*, meaning it treats every command as an expression to be evaluated. For instance:

```
  let accounts = { A2 = 2000, A3 = 2001, B2 = 1492, B3 = 1560,
                   C2 = 971, C3 = 1803, D2 = C2-B2, D3 = C3-B3 }.
  accounts \/
  { E2 = D2*0.33, E3 = D3*0.33 }
```

Excelsior data types include numbers, sets, lists, finite mappings, vectors (useful for representing offsets in spreadsheets), matrices, cell addresses within worksheets, cell ranges, worksheets, formulae or expressions, and equations, as well as (since Excelsior is built on Prolog), arbitrary Prolog terms - and spreadsheets. Typing is dynamic.

### 6.1 Design principles

The reasons for Excelsior's design are explained in Section 3: Excelsior provides modularity by implementing the operators of Section 5. It imitates "scripting languages" such as Perl and Python, by, for example, being a functional language, having dynamic typing, and being concise. It is interactive, and supports trial-and-error assignment of results generated during structure discovery, in the same way that Mathematica and other systems do for computer algebra - [Daly et. al.] compares approaches.

### 6.2 What Excelsior contains

To meet the page limit in these proceedings, I give a short general account rather than, as originally submitted, a detailed specification with examples. Excelsior includes:

**Primitives:** The operators of Section 5, and the notation for sets of equations.

**Variables:** These are defined and assigned with the `let` keyword, as in Section 5.

**Getting formulae:** when editing and patching spreadsheets, we often need to get individual formulae. There is a function `lookup` for this.

**Changing how formulae are interpreted:** When Excelsior reads a spreadsheet, it parses the formulae, converting to expression trees and fixing up relative cell references. Though often the most useful representation, we can ask for others. *Raw* is the formula as a string, for example `R[-33]C+1`. Strings are faster to compare than expression trees, so this is useful when we only need to compare two formulae. *Relative* is like the default representation: the formula as an expression tree, but without fixing up relative references. This is useful in structure discovery, because Excel preserves relative cell references when the user copies a formula from one cell to another. And *fully substituted* is the formula with names replaced by the corresponding item from Excel's named-range list, and with array formulae fixed up.

**Getting style information:** There are functions similar `lookup` for getting stuff such as cell styles and colours, column widths, and named ranges.

**Listing spreadsheets:** The function `show` lists a spreadsheet's equations to terminal or file in human-readable format. This is a handy way to see a spreadsheet's calculations, and to get them printed.

**Loading and saving files:** The function `load` is the Excelsior equivalent of Section 5's `contents_of`. There is also a function for saving spreadsheets to file.

**Running Excel:** The function `excel` runs Excel on a spreadsheet.

**Editing formulae:** The function `replace` acts like a text editor's global search-and-replace, on individual formulae. As an example, I built a small algebraic simplifier which replaces a spreadsheet's formulae by simplified equivalents. I also used it in the accounts-spreadsheet patching trial of Section 9.2, to replace all references to original inputs by the corrected equivalents.

**Spreadsheet grammar rules:** These are described in [Paine, 2004(a)].

**Spreadsheet emulation**: The function `evaluate` is a simple spreadsheet engine. It is far from a complete implementation of the Excel functions, being limited to the basic arithmetic operators and mathematical functions, but it was still useful in running some economic-simulation spreadsheets in a Web-based distance learning system. For a more complete evaluator, I am considering Gnumeric [Gnumeric].

## 7 PARAMETERISATION, GENERALISATION, END-USERS, AND GRAPHICAL USER INTEERFACES

So far, I have written as though each module will do the same thing every time it is used. However, we shall often want to change, for instance, the cells from which modules take their inputs, and the size of the tables they work on. So we must be able to give them *parameters* – to tell a module which cells are its inputs, or how big a table is. This can be done by defining Excelsior functions with such arguments.

I have also written as though Excel users will build spreadsheets by typing Excelsior commands. This is not so. Excelsior can be used directly by those with enough programming experience. I want, however, to provide something for those who know only Excel. To do this, I am working on automatically decomposing existing spreadsheets into modules, then generalising these modules so that they can be parameterised as above. With these, users will be able to build spreadsheets and have them automatically converted into reusable modules. This will need a graphical user interface to display newly proposed modules and to splice them into existing spreadsheets. I shall report this work in a future paper. The source of Excel not being available, Gnumeric [Gnumeric] may be a useful vehicle for experiments.

## 8 EXCELSIOR AND PROLOG

Excelsior is implemented in Prolog [Paine, 2004(b)]. For the reasons stated in Section 6.1, I did not make it resemble Prolog. However, some spreadsheet engineers will be willing to learn Prolog: it is, for example, extremely useful for representing structural relations between spreadsheet modules. Therefore, Excelsior enables its user to drop into the Prolog top-level interpreter, and to call Prolog predicates as Excelsior functions.

## 9 APPLICATIONS

Excelsior is new, but there has been time to try four applications. That of Section 9.4 was requested for evaluation by a large US investment bank and, in part, involves checking a spreadsheet for allocating loans of up to 5 billion dollars. The other three are small trials to see how well Excelsior fitted certain tasks. The first, reverse-engineering a cellular automaton, shows how a huge spreadsheet can be compressed into a tiny listing, which can then be parameterised and compiled to generate variants of this spreadsheet.

### 9.1 Reverse-engineering and simplifying a cellular automaton

This trial used the small-worlds cellular automaton of [Complex Analysis]; [Hand, 2005] explains how such spreadsheets work. Because it lays out 26 successive generations of the cellular automaton down one worksheet, the small-worlds spreadsheet holds a vast amount of repetitive structure, both within each generation's grid, and from one grid to another. Using Excelsior functions for detecting and grouping identical formulae, I could list the spreadsheet very compactly, reducing a 1.8 MB Excel file to a 14K plain text listing containing 17 equations. I was then able to change the state-change formulae in this listing and feed it through the Excelsior compiler, generating a spreadsheet simulating a new cellular automaton. To generate such a new cellular automaton directly in Excel, I would have had to edit about 800 formulae in the original spreadsheet.

As an example, Excelsior listed one batch of repeated formulae as:

```
  Sheet1[ {1} >< { 37..829 by 33 } ] = Sheet1[ HERE, HERE - 33 ]+1
```

This says that the cells in rows 37, 70, ... 829 of column 1 contain the relative formula

```
  R[-33]C+1.
```

### 9.2 Patching end-of-year accounts

A trial suggested by spreadsheet expert Duncan Williamson used Excelsior to edit accounts spreadsheets. These often fail to balance at the end of the accounting year, because some credits or debits are wrong. Accountants fix this by patching with corrections such as "The sales department must be noted to have spent an extra £200". To do these edits directly in Excel, we would have to insert extra rows containing the correction, then update all cell references so they refer to these corrected values rather than the originals. In contrast, it was easy to code the edits in Excelsior, and also to reformat the spreadsheet to the cascaded layout Williamson recommends [Williamson].

### 9.3 Converting spreadsheets to other programming languages

I have experimented with Excelsior to dump spreadsheet-based simulations as equations for reimplementation in other programming languages. These include Java, Fortran, and Lumina's Analytica modelling package.

### 9.4 Comparing and stylechecking an investment bank's credit application and other spreadsheets

A large and well-known US investment bank is evaluating Excelsior for stylechecking and security. One stylechecking program written in Excelsior checks spreadsheets follow the bank's guidelines that only one copy of any formula occurs on each worksheet. Another application concerns a spreadsheet, copies of which businesses fill in when applying for a loan. To check the applicants haven't tampered with formulae in their copy, I have written an Excelsior program for comparing copies against the original. This application is crucial, as the spreadsheet is used for loans of up to 5 billion dollars.

## 10 ACKNOWLEDGEMENTS

To Andreas, Jack and Kostas, at the Excelsior café, Cowley Road. Something that's excelsior does more then excel.